\documentclass[aps,pra,floatfix,twocolumn,showpacs,amsmath,amssymb,letterpaper]{revtex4-2}
\usepackage{graphicx}
\usepackage{array}
\usepackage{bigstrut}
\usepackage{longtable}
\usepackage{rotating}
\usepackage{diagbox}
\usepackage[mode=text]{siunitx}
\usepackage{bm}
\usepackage{float}
\usepackage{lipsum}
\usepackage{warpcol}
\usepackage{color}
\usepackage{amsthm,amsmath}
\usepackage{mathrsfs}
\usepackage{appendix}
\usepackage{multirow}
\usepackage{threeparttable}
\usepackage{soul}
\usepackage[colorlinks=true, linkcolor=blue, citecolor=blue, urlcolor=blue]{hyperref}
\usepackage{nameref}
\usepackage{csquotes}
\usepackage{tabularx}
\usepackage{booktabs} 
\usepackage{dcolumn}
\allowdisplaybreaks

\newcolumntype{d}[1]{D{.}{.}{#1}}

\begin{document}



\title{Precise \emph{ab initio} calculations of $^4\mathrm{He}$($1snp\,^3P_J$) fine structure of high Rydberg states }



\author{Hao Fang$^{1}$}
\thanks{These authors contributed equally to this work.}
\author{Jing Chi$^{1, 2}$}
\thanks{These authors contributed equally to this work.}
\author{Xiao-Qiu Qi$^3$}
\author{Yong-Hui Zhang$^{1}$}
\email{yhzhang@wipm.ac.cn}
\author{Li-Yan Tang$^{1}$}
\email{lytang@apm.ac.cn}
\author{Ting-Yun Shi$^{1}$}

\affiliation{
	$^{1}$
	Innovation Academy for Precision Measurement Science and Technology,
	Chinese Academy of Sciences, Wuhan 430071, China
}

\affiliation{
	$^{2}$University of Chinese Academy of Sciences, Beijing 100049, China
}

\affiliation{
	$^{3}$Department of Physics, Zhejiang Sci-Tech University, Hangzhou 310018, China
}

\date{\today}

\begin{abstract}

	High-precision measurements of the fine-structure splittings in helium high Rydberg states have been reported, yet corresponding ab initio benchmarks for direct comparison remain unavailable. In this work, we extend the correlated B-spline basis function (C-BSBF) method to calculate the fine-structure splittings of high Rydberg states in $^4$He. The calculations include the $m\alpha^4$- and $m\alpha^5$-order contributions, the singlet–triplet mixing effect, and estimated spin-dependent $m\alpha^6$-order corrections obtained using a $1/n^3$ scaling approximation. High-precision ab initio results are obtained for principal quantum numbers $n=24$–37 with kilohertz-level accuracy and further extended to $n=45$–51 by extrapolation and fitting. The theoretical results show excellent agreement with quantum-defect theory (QDT) predictions and allow direct comparison with experimental measurements. Additionally, the discrepancy observed at $n=34$ is expected to be clarified with improved experimental precision.
\end{abstract}

\keywords{Helium atom, fine structure, B-spline basis, Breit-Pauli Hamiltonian, Rydberg states, QED corrections}
\pacs{31.30.J-, 31.15.-p, 31.15.ac}

\maketitle

\section{Introduction}


The helium fine structure has attracted significant theoretical and experimental interest ever since Schwartz proposed the intervals of the $1s2p\,^3P_J$ states as a sensitive probe of the fine-structure constant $\alpha$~\cite{schwartz1964}. The $J=2\rightarrow0$ interval has been measured with an accuracy of 130 Hz, which, if matched by complete $m\alpha^8$-order QED calculations, could enable a determination of the fine-structure constant at the 2-ppb level \cite{zheng2017}. Nevertheless, the required theoretical calculations appear to be too complicated to be accomplished
in the near future~\cite{Pachucki2017}. For other two intervals of $J=2\rightarrow1$ and $J=1\rightarrow0$, the current measurements have achieved a precision on the order of tens of Hz~\cite{Kato2018, Heydarizadmotlagh2024}. However, these results deviate from theoretical predictions~\cite{pachucki2010} by 1.5 and 1.6 $\sigma$, respectively. Further advances in theoretical precision are mainly hindered by uncertainties in higher-order QED corrections~\cite{pachucki2006, pachucki2010}. In addition, for the Rydberg fine-structure intervals with principle quantum numbers $n$ below 10, high-precision calculations and measurements have been performed in order to test the long-range interactions between the Rydberg electron and the ionic core~\cite{hessels1992, claytor1995, storry1995, stevens1999}.

For higher values of $n$, measurements of the fine structure in helium are of great importance for various applications, including hybrid cavity quantum electrodynamics and quantum information processing~\cite{rabl2006, hogan2012}, as well as for studies of F\"{o}rster resonance energy transfer in collisions with polar ground-state molecules~\cite{zhelyazkova2017}. Significant progress has been made in measuring the fine structures of these higher Rydberg states~\cite{Deller2018}. Using microwave spectroscopy, Deller and Hogan observed the fine-structure intervals for $n=$ 34 to 36, and also resolved the large intervals of more highly excited states with $n$ between 45 and 51, achieving transition-frequency uncertainties of 15--40 kHz~\cite{Deller2018}. Most of the measured intervals agreed well with the quantum defect results~\cite{Drake1999}, but a significant discrepancy was found for $n=$ 34~\cite{Deller2018}. Given that quantum defect theory (QDT) is not an exact fundamental theory~\cite{drake1994}, and that relativistic and QED corrections for higher Rydberg states scale as $1/n^3$, higher-order effects can be neglected. Consequently, high-precision \emph{ab initio} calculations of the fine structures in helium Rydberg states are both practically feasible and theoretically essential.

Theoretically, benchmark \emph{ab initio} data remain concentrated on low-lying states (roughly $n\!\le\!10$ with $L\!\le\!7$)~\cite{Drake1992, drake1998, Aznabaev2018} while analyses of higher $n$ states typically rely on QDT, since conventional variational expansions rapidly lose accuracy with increasing $n$ owing to the large radial-scale separation between the core and Rydberg electrons. Recently, by employing “triple” Hylleraas basis sets that simultaneously resolve short-range correlation and long-range Rydberg behavior, Bondy \emph{et al.}~\cite{Bondy2025} overcame this limitation and reported a kHz-level determination of the ionization energy for the $24 \, ^1P_1$ state. This pioneering high-accuracy \emph{ab initio} calculation in the high-$n$ regime demonstrated that a systematic extension of precision theoretical methods to Rydberg states is now feasible. The Rydberg states investigated experimentally~\cite{Deller2018, clausen2021ionization, clausen2025metrology, clausen2025ionization} typically lie at much higher principal quantum numbers. The above approach employing “triple” Hylleraas basis sets has only very recently enabled ab initio calculations to be extended to as high as $n=$ 35~\cite{drake2025}; however, their work has mainly focused on the ionization energy.


In contrast to multiple-scale Hylleraas basis, we developed the correlated B-spline basis-function (C-BSBF) method to enable high-precision calculations for both low-lying and Rydberg states of helium~\cite{yang2017application,yang2019application,fang2023application,fang2024sub}. The C-BSBFs inherently accommodate the multiscale character of Rydberg helium. Accurate energies and polarizabilities have previously been reported for states up to $n=$ 10 ~\cite{yang2017application,yang2019application,fang2024sub}, and more recently, nonrelativistic energies of the Rydberg $1snp \,^{1,3}P$ states up to $n=$ 27~\cite{chi2025accurate} were obtained with an accuracy better than fourteen significant digits.
B-splines posses the advantages of completeness and linear independence~\cite{Bachau2001, Charlotte2008}. Building on these advantages, the developed C-BSBF method explicitly incorporated
the interelectronic coordinate $r_{12}$, allowing it to simultaneously describe short-range electron-electron cusps and long-range asymptotic behavior within a unified basis, thus eliminating the nonlinear-parameter optimization over different radial sectors required in Hylleraas calculations~\cite{Bondy2025}.

In this work, the C-BSBF method is extended to compute $1snp\,^3P_J$ fine-structure intervals of high Rydberg states in $^4\mathrm{He}$, enabling direct comparison of ab initio results with experimental data. The leading relativistic corrections with nuclear recoil, the anomalous magnetic moment, and singlet-triplet mixing effects are taken into account. Ab initio results are presented for $n=24-37$, while data for $n=45-51$ obtained by the $1/n$ asymptotic expansion are also reported.
%
The inverse fine-structure constant, the electron mass, the nuclear mass of $^4\mathrm{He}$, and the Rydberg frequency are taken as $\alpha^{-1} = \num{137.035 999 177(21)}$, $m_e=$ 1, $M_0 = \num{7294.299 541 71(17)} \, m_e$, and $c R_y = \num{3.289 841 960 2500(36)} \times 10^{9}$ MHz~\cite{CODATA2022}, respectively.

\section{Theoretical Method}   
\begin{figure}
	\centering
	\includegraphics[scale=0.205]{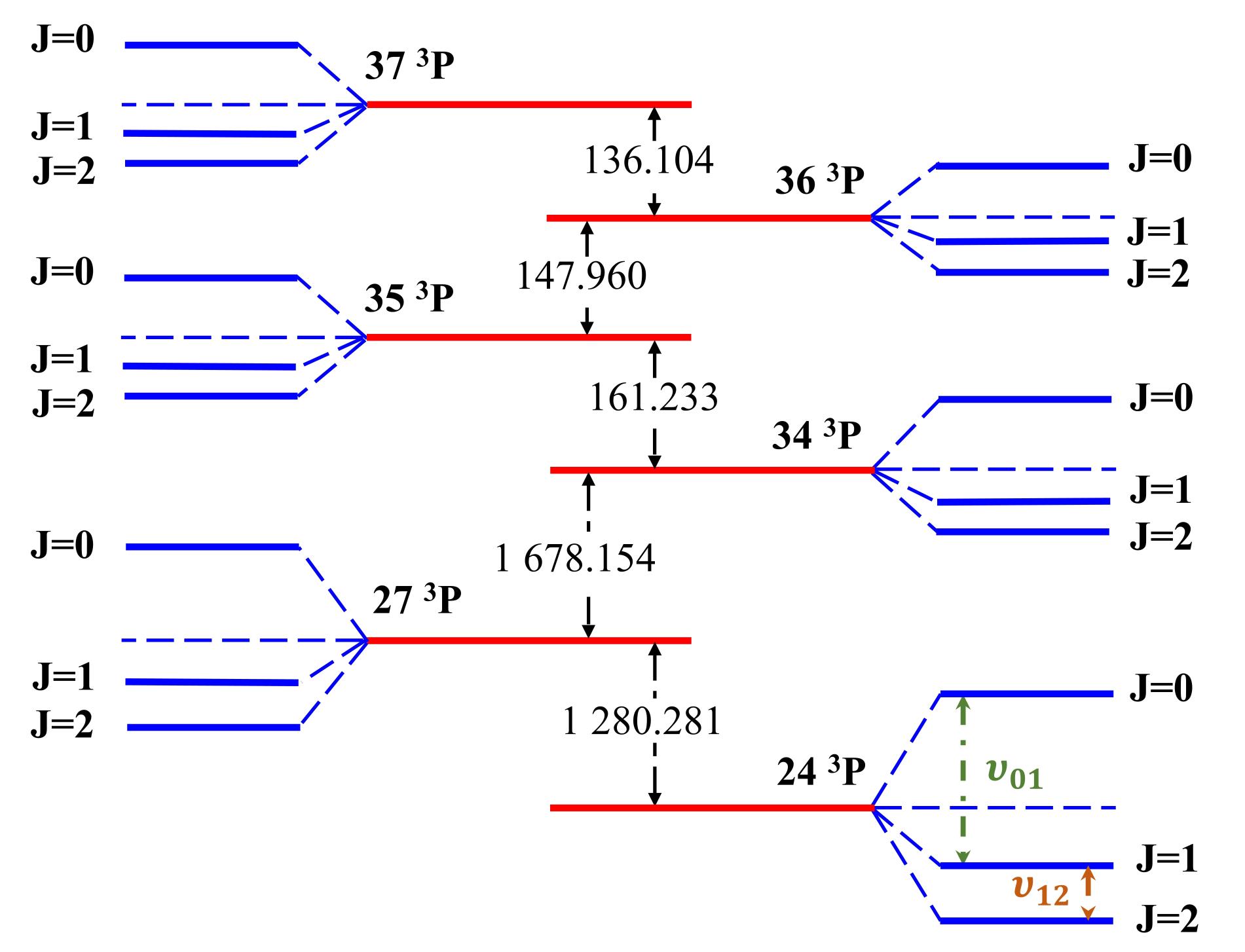}
	\caption{Fine-structure energy levels (not drawn to scale) for the $n^3P_J\,(J=0,1,2)$ states of $^4\mathrm{He}$, in GHz.}
	\label{fig:energy_level}
\end{figure}

Within the nonrelativistic quantum electrodynamics framework, the fine structure splittings of the $n\,^3P_J$ states are obtained by first solving the nonrelativistic Schrödinger equation to determine unperturbed (zeroth-order) energies and wave functions. The fine-structure effects are then introduced perturbatively through the Breit-Pauli Hamiltonian~\cite{bethe1957, Drake1992}. The fine-structure energy-level diagram of the higher Rydberg states $n\,^3P_J\,(J=0, 1, 2)$ with $n=$ 24, 27, 34, 35, 36, and 37 in $^4$He is displayed in Fig.~\ref{fig:energy_level}.

The nonrelativistic Hamiltonian of $^4$He in atomic units takes the form
\begin{equation}
	H=\frac{p_1^2+p_2^2}{2 \mu_e}
	+\frac{\mathbf{p}_1 \cdot \mathbf{p}_2}{M_0}
	-\frac{Z}{r_1}
	-\frac{Z}{r_2}
	+\frac{1}{r_{12}},
\end{equation}
where $r_i$ is the coordinate of the $i$-th electron to the nucleus, $r_{12}=|\mathbf r_1-\mathbf r_2|$ is the interelectronic coordinate, $M_0$ is the nuclear mass of $^4$He, $\mu_e = m_e M_0/(m_e+M_0)$ is the reduced electron mass, and the nuclear charge $Z = 2$. The $\mathbf{p}_1 \cdot \mathbf{p}_2/M_0$ term is the mass-polarization contribution due to nuclear motion in the center-of-mass frame.

The two-electron wave functions are expanded in C-BSBFs~\cite{fang2023application},
\begin{equation}
	\phi_{ij,c,\ell_1\ell_2}
	=\mathcal{A}\!\left[
		r_{12}^{c}\,
		B_{i}^{k}(r_{1})\,B_{j}^{k}(r_{2})\,
		\mathcal{Y}_{\ell_{1}\ell_{2}}^{LM}
		(\hat{\mathbf r}_{1},\hat{\mathbf r}_{2})
		\right],
\end{equation}
where $\mathcal{A}$ antisymmetrizes the electrons, $c=0$ and $1$, $N$ is the number of B-splines confined in a cavity of radius $R_0$, $B_i^{k}(r)$ is the $i$-th B-spline of order $k$~\cite{Bachau2001}, and $\mathcal{Y}_{\ell_{1}\ell_{2}}^{LM}$ are coupled spherical harmonics~\cite{brinkAngularMomentum1994,drakeAngularIntegralsRadial1978}.
Orbital angular momenta satisfy $\ell_{1},\ell_{2}\le\ell_{\max}$.



The leading contribution to the fine-structure splittings originates from the spin-dependent Breit-Pauli interaction~\cite{bethe1957} of order $m\alpha^4$, since the spin-independent part
contributes equally to all fine structure levels. The effective spin-dependent Breit-Pauli Hamiltonian can be expressed as
\begin{equation}
	H_{\mathrm{fs}}^{(4)}
	=H_{\mathrm{so}}
	+H_{\mathrm{soo}}
	+H_{\mathrm{ss}}
	+\Delta ,
\end{equation}
with
\begin{align}
	H_{\mathrm{so}}      & =
	\frac{Z \alpha^2}{2}\!\left[
		\frac{\mathbf r_1}{r_1^{3}}\!\times\!\mathbf p_1\!\cdot\!\mathbf s_1
		+\frac{\mathbf r_2}{r_2^{3}}\!\times\!\mathbf p_2\!\cdot\!\mathbf s_2
	\right], \nonumber        \\
	H_{\mathrm{soo}}     & =
	\frac{\alpha^2}{2r_{12}^{3}}\!\Bigl[
		\mathbf r_{12}\!\times\!\mathbf p_{2}\!\cdot\!(2\mathbf s_1+\mathbf s_2)
		-\mathbf r_{12}\!\times\!\mathbf p_{1}\!\cdot\!(2\mathbf s_2+\mathbf s_1)
	\Bigr], \nonumber         \\
	H_{\mathrm{ss}}      & =
	\alpha^2 \left[ \frac{\mathbf s_1\!\cdot\!\mathbf s_2}{r_{12}^{3}}
		-\frac{3(\mathbf s_1\!\cdot\!\mathbf r_{12})
			(\mathbf s_2\!\cdot\!\mathbf r_{12})}{r_{12}^{5}}
	\right], \nonumber        \\
	\Delta_{\mathrm{so}} & =
	Z \alpha^2\!\left(
	\frac{\mathbf r_1}{r_1^{3}}\!\times\!\mathbf {p}_{2}\!\cdot\!\mathbf s_1
	+\frac{\mathbf r_2}{r_2^{3}}\!\times\!\mathbf {p}_{1}\!\cdot\!\mathbf s_2
	\right),
\end{align}
where $H_{\mathrm{so}}$ is the spin-orbit interaction, $H_{\mathrm{soo}}$ is the spin-other-orbit interaction, $H_{\mathrm{ss}}$ is  the spin-spin interaction, $\Delta=\frac{m_e}{M_0}\bigl(\Delta_{\mathrm{so}}+2H_{\mathrm{so}}\bigr)$ is the Stone term arising in the center-of-mass frame~\cite{stone1961nuclear,stone1963nuclear}, and $\mathbf s_{1,2}$ are Pauli spin matrices.


In addition, the contribution of order $m\alpha^5$ arises from the electron anomalous magnetic moment corrections to Breit-Pauli operators, and the effective Hamiltonian has the form of
%
%
%
\begin{equation}
	H_{\mathrm{fs}}^{(5)}
	=\frac{\alpha}{\pi}\!\left[
		H_{\mathrm{so}}
		+\frac{2}{3}H_{\mathrm{soo}}\,
		\delta_{SS'}
		+H_{\mathrm{ss}}
		\right],
\end{equation}
where $\delta_{S,S'}$ nulls the single-triplet mixing part of $H_{\mathrm{soo}}$~\cite{araki1959triplet,hambro1972secondorder,lewis1978secondorder}.


The singlet–triplet mixing correction to the $n\,^3P_1$ energy level, induced by the spin-dependent Breit–Pauli operators, is of order $m\alpha^6$. This effect includes both fine- and hyperfine-induced mixing, of which only the fine component is relevant for $^4$He.
For higher Rydberg states (larger $n$), the energy interval between the $n\,^3P_1$ and $n\,^1P_1$ levels becomes progressively smaller; consequently singlet–triplet mixing grows in significance and must be evaluated with high accuracy. The singlet–triplet contribution is obtained by exact diagonalization of an effective $2\times 2$ Hamiltonian~\cite{drake1979,Drake1992,drake1988}, where the mixing correction $\Delta_{\mathrm{mix}}$ is defined as the difference between the eigenvalue and the corresponding diagonal element:
\begin{equation}
	H_{\mathrm{eff}}=
	\begin{pmatrix}
		E_{\mathrm{dia}}(n \, ^3P_1) & E_{\mathrm{off}}              \\
		E_{\mathrm{off}}             & E_{\mathrm{dia}}(n \, ^1P_1)
	\end{pmatrix} \, ,
\end{equation}
where $E_{\mathrm{dia}}(n \, ^3P_1)$ and $E_{\mathrm{dia}}(n \, ^1P_1)$ correspond to the energies of the triplet and singlet states, respectively, and $E_{\mathrm{off}}$ denotes the off-diagonal matrix element responsible for fine mixing, which can be expressed as
\begin{equation}
	\begin{aligned}
		E_\mathrm{off} & =\langle n \, ^3P_1 \left| H_{\mathrm{mix}} \right| n \, ^1P_1 \rangle       \\
		               & =\langle n \, ^1P_1 \left| H_{\mathrm{mix}} \right| n \, ^3P_1 \rangle \, ,
	\end{aligned}
\end{equation}
here
\begin{equation}
	H_{\mathrm{mix}}=H_{\mathrm{so}}+H_{\mathrm{soo}} \, . 
\end{equation}

In the $|LSJM_J\rangle$ basis, the coupled state can be expressed as
\begin{equation}
	|LSJM_J\rangle
	= \sum_{M_LM_S}
	|LSM_LM_S\rangle
	\langle LSM_LM_S|LSJM_J\rangle,
\end{equation}
where $|LSM_LM_S\rangle$ denotes the uncoupled LS basis state, in which the orbital angular momentum $\mathbf L$ and total spin $\mathbf S$ are separately quantized before coupling to the total angular momentum $\mathbf J = \mathbf L + \mathbf S$.
In this representation, the expectation value of a spin-dependent operator $\hat X$ factorizes as
\begin{equation}
	\langle \hat X \rangle_J
	= \Lambda(J)\,\langle \hat X \rangle,\label{pj}
\end{equation}
where $\langle \hat X \rangle$ denotes the expectation value evaluated in the uncoupled LS basis. The angular-momentum coupling coefficients are given by $\Lambda_J = \{-1/3, -1/6, 1/6\}$ for $H_{\mathrm{so}}$ and $H_{\mathrm{soo}}$, and $\{1/3, -1/6, 1/30\}$ for $H_{\mathrm{ss}}$ at $J=0,1,2$, respectively.

The fine-structure splittings follow as
%
%
\begin{equation}
	\begin{aligned}
		\nu_{01} =  & -\frac{1}{6}\left[\left\langle H_{\mathrm{so}}\right\rangle+\left\langle H_{\mathrm{soo}}\right\rangle+\langle\Delta\rangle\right]+\frac{1}{2}\left\langle H_{\mathrm{ss}}\right\rangle                             \\
		            & -\frac{1}{6}\left[\dfrac{\alpha}{\pi}\left\langle H_{\mathrm{so}}\right\rangle+\dfrac{2\alpha}{3\pi}\left\langle H_{\mathrm{soo}}\right\rangle\right]+\dfrac{\alpha}{2\pi}\left\langle H_{\mathrm{ss}}\right\rangle \\
		            & -\Delta_{\mathrm{mix}},                                                                                                                                                                                             \\
		\nu_{12}  = & -\frac{1}{3}\left[\left\langle H_{\mathrm{so}}\right\rangle+\left\langle H_{\mathrm{soo}}\right\rangle+\langle\Delta\rangle\right]-\frac{1}{5}\left\langle H_{\mathrm{ss}}\right\rangle ,                           \\
		            & -\frac{1}{3}\left[\dfrac{\alpha}{\pi}\left\langle H_{\mathrm{so}}\right\rangle+\dfrac{2\alpha}{3\pi}\left\langle H_{\mathrm{soo}}\right\rangle\right]-\dfrac{\alpha}{5\pi}\left\langle H_{\mathrm{ss}}\right\rangle \\
		            & +\Delta_{\mathrm{mix}}.
	\end{aligned}
\end{equation}
In practice, one evaluates the combined interval $\nu_{02}=\nu_{01}+\nu_{12}$, for which singlet--triplet mixing cancels exactly~\cite{pachucki2009reexamination}, whereas this mixing has a significant effect on the individual intervals $\nu_{01}$ and $\nu_{12}$.
%



\section{Results and Discussions}

In our previous work~\cite{chi2025accurate}, a cavity radius of $R_0=$ 2400 a.u. allowed accurate calculations of nonrelativistic energies up to $n=$ 27. In the present study, the extension to higher Rydberg states inevitably requires a larger cavity radius, as the electronic radial distribution at $n \sim 37$ spans much larger distances. Through semi-empirical parameters adjusting, the cavity radius is currently chosen to be $R_0=$ 4000 a.u.. Increasing the cavity size, however, imposes more stringent demands on numerical integration. To ensure computed accuracy and stability across the entire radial interval, numerically improved Laplace expansions and an efficient adaptive integration scheme were employed.

Calculations were performed using B-splines of sizes $N=$ 80, 90, and 100 for $\ell_{\max}=4$, while an additional calculation with $N=100$ and $\ell_{\max}=5$ was carried out to evaluate the influence of the $\ell_{\max}$ truncation. 
The corresponding basis dimensions reached up to 80 000 and 100 000 for $\ell_{\max}=4$ and 5, respectively, and full diagonalization was performed using a self-developed quadruple-precision parallel program.
The convergence uncertainty was defined as the larger of the two sources of numerical deviation: (i) the maximum difference between the extrapolated value and the results from the three largest basis sets ($N=80$, $90$, and $100$ at $\ell_{\max}=4$), and (ii) the change in the result when $\ell_{\max}$ was increased from 4 to 5.

\begin{table*}[ht]
	\footnotesize
	\caption{\label{tab:nonrel-energy-operator} Nonrelativistic energies (in a.u.) of the $n\,^3P$ states and expectation values (in $10^{-5}\alpha^2$ a.u.) of spin-dependent Breit operators for the $n\,^3P_1$ states of $^4$He ($n=24$–$37$). For each principal quantum number, the first line presents the current results for $^4$He, the second line lists Hylleraas reference data from Ref.~\cite{drake2025}. The calculations adopt node parameter $\tau = 0.00228$, spline order $k=15$, $\ell_{\max}=4$, and box size $R_0=4000\,\mathrm{a.u.}$.  The notation $\langle X \rangle_1^+$ indicates the expectation value including finite nuclear mass effects. The numbers in parentheses are the convergence uncertainties.}
	\begin{threeparttable}
		\begin{tabular*}{\textwidth}{@{\extracolsep{\fill}} c d{4.22} d{4.14} d{4.15} d{4.16}  d{4.20}}
			\toprule
			$n$ & \multicolumn{1}{c}{$^{4}\text{He}$} & \multicolumn{1}{c}{$\langle H_{\text{so}}\rangle_1^+$}    & \multicolumn{1}{c}{$\langle H_{\text{soo}}\rangle_1^+$}   & \multicolumn{1}{c}{$\langle H_{\text{ss}}\rangle_1^+$}    & \multicolumn{1}{c}{$\langle \Delta_{\text{so}}\rangle_1^+$}       \\
			\midrule
			24 & -2.000\,598\,750\,318\,682\,937(1)   & -1.854\,878\,538\,840\,1(2)    &  2.675\,672\,79(1)           & -1.087\,838\,65(1)            & 5.428\,180\,049\,382\,4(4) \\
			&  &                                        -1.854\,878\,538\,847(9)       &  2.675\,672\,783\,442(7)     & -1.087\,838\,649\,784\,7(15)  & 5.428\,180\,049\,6(18)     \\
			25 & -2.000\,530\,131\,464\,088\,097(6)   & -1.640\,515\,298\,743\,1(4)    &  2.366\,412\,90(1)           & -0.962\,070\,75(1) 	          & 4.800\,863\,560\,988(1)    \\
			&  &                                        -1.640\,515\,298\,741(9)       &  2.366\,412\,898\,439(9)     & -0.962\,070\,744\,534\,1(32)  & 4.800\,863\,561\,0(5)      \\
			26 & -2.000\,469\,296\,887\,281\,00(2)    & -1.457\,950\,682\,678\,1(1)    &  2.103\,035\,55(1)           & -0.854\,967\,89(1) 	          & 4.266\,602\,516\,982\,2(2) \\
			&  &                                        -1.457\,950\,682\,662(19)      &  2.103\,0355\,444\,51(24)    & -0.854\,967\,883\,837(6)      & 4.266\,602\,517\,38(20)    \\
			27 & -2.000\,415\,112\,195\,609\,93(4)    & -1.301\,500\,491\,618(2)       &  1.877\,337\,53(1)           & -0.763\,191\,82(1) 	          & 3.808\,762\,819\,451(6)    \\
			&  &                                        -1.301\,500\,491\,637\,4(16)   &  1.877\,337\,528\,619(9)     & -0.763\,191\,821\,791\,0(8)   & 3.808\,762\,822\,8(9)      \\
			28 & -2.000\,366\,642\,353\,614\,61(8)    & -1.166\,657\,507\,730(4)       &  1.682\,814\,55(1)           & -0.684\,095\,94(1) 	          & 3.414\,154\,626\,81(1)     \\
			&  &                                        -1.166\,657\,507\,732(7)       &  1.682\,814\,549\,259(22)    & -0.684\,095\,937\,963(8)      & 3.414\,154\,625\,2(4)      \\
			29 & -2.000\,323\,111\,072\,696\,5(2)     & -1.049\,816\,704\,402(8)       &  1.514\,264\,48(1)           & -0.615\,563\,66(1) 	          & 3.072\,228\,231\,26(2)     \\
			&  &                                        -1.049\,816\,704\,323(11)      &  1.514\,264\,474\,503\,4(27) & -0.615\,563\,661\,817\,5(11)  & 3.072\,228\,04(8)          \\
			30 & -2.000\,283\,869\,534\,359\,9(8)     & -0.948\,071\,397\,66(2)        &  1.367\,493\,06(1)           & -0.555\,888\,64(1) 	          & 2.774\,477\,530\,87(6)     \\
			&  &                                        -0.948\,071\,397\,669(10)      &  1.367\,493\,061\,54(7)      & -0.555\,888\,638\,049(24)     & 2.774\,477\,524(11)        \\
			31 & -2.000\,248\,372\,073\,557(2)        & -0.859\,060\,382\,47(6)        &  1.239\,093\,30(1)           & -0.503\,684\,89(1) 	          & 2.513\,992\,698\,3(2)      \\
			&  &                                        -0.859\,060\,382\,470\,2(30)   &  1.239\,093\,296\,767\,2(37) & -0.503\,684\,883\,671\,5(11)  & 2.513\,992\,700\,13(9)     \\
			32 & -2.000\,216\,157\,106\,699(7)        & -0.780\,852\,144\,9(2)         &  1.126\,278\,27(1)           & -0.457\,818\,75(1) 	          & 2.285\,121\,345\,4(6)      \\
			&  &                                        -0.780\,852\,144\,890\,2(37)   &  1.126\,278\,262\,651(7)     & -0.457\,818\,747\,772(7)      & 2.285\,121\,24(5)          \\
			33 & -2.000\,186\,832\,050\,30(2)         & -0.711\,856\,333\,0(6)         &  1.026\,753\,36(1)           & -0.417\,356\,90(1) 	          & 2.083\,209\,455(2)         \\
			&  &                                        -0.711\,856\,332\,881(10)      &  1.026\,753\,359\,756\,4(15) & -0.417\,356\,898\,580\,8(5)   & 2.083\,209\,31(11)         \\
			34 & -2.000\,160\,061\,303\,70(7)         & -0.650\,755\,474(2)            &  0.938\,617\,76(1)           & -0.381\,526\,21(1) 	          & 1.904\,401\,561(6)         \\
			&  &                                        -0.650\,755\,4737\,1(16)       &  0.938\,617\,756\,3(15)      & -0.381\,526\,212\,0(8)        & 1.904\,402\,1(4)           \\
			35 & -2.000\,135\,556\,604\,6(2)          & -0.596\,451\,875(6)            &  0.860\,287\,75(1)           & -0.349\,682\,58(1) 	          & 1.745\,485\,36(2)          \\
			&  &                                        -0.596\,451\,870\,7(25)        &  0.860\,287\,753\,1(6)       & -0.349\,682\,583\,772\,33(6)  & 1.745\,482(4)              \\
			36 & -2.000\,113\,069\,236\,5(7)          & -0.548\,026\,01(2)             &  0.790\,436\,73(3)           & -0.321\,286\,49(1) 	          & 1.603\,769\,93(6)          \\
			37 & -2.000\,092\,383\,692(8)             & -0.504\,703\,7(8)              &  0.727\,948(1)               & -0.295\,883\,8(6) 		      & 1.476\,989(2)              \\
			\bottomrule
		\end{tabular*}
	\end{threeparttable}
\end{table*}
At first, energies of the $n\,^3P$ series were calculated for both infinite nuclear mass of helium and $^4$He. For the case of $n=24$ with infinite nuclear mass, the present C-BSBF calculation yields an energy of $\num{-2.000873014566616659(2)}$ a.u., which agrees with the Hylleraas result~\cite{Bondy2025} to eighteen significant digits, confirming the high accuracy of the present approach. Nonrelativistic energies of the $n\,^3P$ states with $n=24-37$ for $^4$He are listed in Table~\ref{tab:nonrel-energy-operator}. The calculated energies are found to have converged to at least 13 significant digits.
All these states were obtained from a single diagonalization without any state-dependent parameter optimization, demonstrating the numerical stability and computational efficiency of the C-BSBF method in describing higher Rydberg states. In addition, we analyzed the effect of the uncertainty in the nuclear mass of $^4$He, $M_0$~\cite{CODATA2022}, on the nonrelativistic energies.
This uncertainty leads to variations at the 14th–15th decimal places, which do not affect the fine-structure splittings calculated in the following sections and are therefore neglected in the present work.

Using Eq.~(\ref{pj}), the contributions of the spin-dependent operators to the different $n\,^3P_J$ states are obtained.
The results for the $n\,^3P_1$ state are listed in Table~\ref{tab:nonrel-energy-operator}, where the notation $\left\langle X \right\rangle_1^+$ denotes the expectation value including the finite nuclear mass correction. These values are required for evaluating the $m\alpha^4$- and $m\alpha^5$-order contributions to the fine-structure splittings.
The numerical accuracy of these values ensures a precision at the kHz level in the fine-structure splitting calculations. Table~\ref{tab:nonrel-energy-operator} also compares the present C-BSBF results with the Hylleraas data~\cite{drake2025}, where the latter were obtained by combining the infinite-nuclear-mass matrix elements with the corresponding mass-polarization corrections. The two sets of matrix elements differ only beyond the sixth significant figure, resulting in differences in the fine-structure splittings below 10 Hz—well below the current experimental accuracy and therefore negligible.

\begin{table*}[ht]
	\caption{Breakdown of the fine-structure intervals $\nu_{12}$ and $\nu_{02}$ (in MHz) of the $n\,^3P_J$ states of $^4$He for $n=24$–$37$. The entries are labeled as follows: “$m\alpha^4$” is the relativistic correction including finite nuclear mass effects; “$m\alpha^5$” is anomalous magnetic moment correction; “$\Delta_{\mathrm{mix}}$” is the singlet-triplet mixing term; “Total” is the total fine-structure splittings. Present results are compared with Hylleraas values~\cite{drake2025}, QDT, and experimental data~\cite{Deller2018}. \label{tab:fine-structure_v12}}
	\begin{threeparttable}
		\begin{tabular*}{\textwidth}{@{\extracolsep{\fill}} c  S[table-format=3.7] S[table-format=4.8] S[table-format=4.8] S[table-format=2.6] S[table-format=4.8] S[table-format=4.7] d{2.6}}
			\toprule
			\multirow{1}{*}{$n$} & \multicolumn{1}{c}{$m\alpha^4$}   & \multicolumn{1}{c}{$m\alpha^5$}  & \multicolumn{1}{c}{$\Delta_{\mathrm{mix}}$}    & \multicolumn{1}{c}{Total} &\multicolumn{1}{c}{Hylleraas~\cite{drake2025}} &\multicolumn{1}{c}{QDT}   & \multicolumn{1}{c}{Exp.~\cite{Deller2018}} \\ \midrule
			\multicolumn{7}{c}{$\nu_{12}$} \\
			24        &  1.179 556    & -0.011 783    & -0.002 266    & 1.165 5(38)        & 1.166 1(16)    & 1.165 6         &             \\
			25        &  1.043 168    & -0.010 421    & -0.002 005    & 1.030 7(33)        & 1.031 5(16)    & 1.030 8         &             \\
			26        &  0.927 023    & -0.009 262    & -0.001 782    & 0.916 0(30)        & 0.916 8(14)    & 0.916 0         &             \\
			27        &  0.827 501    & -0.008 268    & -0.001 591    & 0.817 6(26)        & 0.818 5(14)    & 0.817 7         &             \\
			28        &  0.741 732    & -0.007 411    & -0.001 426    & 0.732 9(24)        & 0.733 9(14)    & 0.732 9         &             \\
			29        &  0.667 419    & -0.006 669    & -0.001 283    & 0.659 5(21)        & 0.660 5(14)    & 0.659 5         &             \\
			30        &  0.602 711    & -0.006 023    & -0.001 159    & 0.595 5(19)        & 0.596 6(14)    & 0.595 5         &             \\
			31        &  0.546 106    & -0.005 457    & -0.001 051    & 0.539 6(17)        & 0.540 8(13)    & 0.539 6         &             \\
			32        &  0.496 373    & -0.004 960    & -0.000 955    & 0.490 5(16)        & 0.491 7(13)    & 0.490 5         &             \\
			33        &  0.452 500    & -0.004 522    & -0.000 871    & 0.447 1(14)        & 0.448 3(13)    & 0.447 1         &             \\
			34        &  0.413 650    & -0.004 134    & -0.000 796    & 0.408 7(13)        & 0.410 0(13)    & 0.408 7         & 0.397       \\
			35        &  0.379 123    & -0.003 789    & -0.000 730    & 0.374 6(12)        & 0.375 9(13)    & 0.374 6         & 0.366       \\
			36        &  0.348 334    & -0.003 481    & -0.000 670    & 0.344 2(11)        &                & 0.344 2         & 0.340       \\
			37        &  0.320 791    & -0.003 206    & -0.000 617    & 0.317 0(10)        &                & 0.317 0         &             \\
			\multicolumn{7}{c}{} \\
			\multicolumn{7}{c}{$\nu_{02}$} \\
			24        & 15.490 868    & 0.014 205     &               & 15.505 1(47)      & 15.501 3(16)   & 15.505 1        &             \\
			25        & 13.699 902    & 0.012 562     &               & 13.712 5(42)      & 13.709 2(16)   & 13.712 5        &             \\
			26        & 12.174 736    & 0.011 163     &               & 12.185 9(37)      & 12.183 1(14)   & 12.185 9        &             \\
			27        & 10.867 829    & 0.009 964     &               & 10.877 8(33)      & 10.875 3(14)   & 10.877 8        &             \\
			28        & 9.741 493     & 0.008 931     &               & 9.750 4(30)       & 9.748 4(14)    & 9.750 4         &             \\
			29        & 8.765 586     & 0.008 036     &               & 8.773 6(27)       & 8.771 8(14)    & 8.773 6         &             \\
			30        & 7.915 809     & 0.007 257     &               & 7.923 1(24)       & 7.921 5(14)    & 7.923 1         &             \\
			31        & 7.172 425     & 0.006 575     &               & 7.179 0(22)       & 7.177 7(13)    & 7.179 0         &             \\
			32        & 6.519 290     & 0.005 976     &               & 6.525 3(20)       & 6.524 1(13)    & 6.525 3         &             \\
			33        & 5.943 113     & 0.005 448     &               & 5.948 6(18)       & 5.947 6(13)    & 5.948 6         &             \\
			34        & 5.432 884     & 0.004 980     &               & 5.437 9(17)       & 5.437 1(13)    & 5.437 9         & 5.387       \\
			35        & 4.979 431     & 0.004 564     &               & 4.984 0(15)       & 4.983 3(13)    & 4.984 0         & 5.014       \\
			36        & 4.575 072     & 0.004 194     &               & 4.579 3(14)       &                & 4.579 3         & 4.643       \\
			37        & 4.213 34      & 0.003 862     &               & 4.217 2(13)       &                & 4.217 2         &             \\
			\bottomrule
		\end{tabular*}
	\end{threeparttable}
\end{table*}
The $m\alpha^4$- and $m\alpha^5$-order contributions to the fine-structure intervals $\nu_{12}$ and $\nu_{02}$ in $^4$He, obtained using the C-BSBF method, are summarized in Table~\ref{tab:fine-structure_v12}.
For the $J=1$ level, the singlet–triplet mixing term contributes at order $m\alpha^6$, while it has no effect for $J=0$ or $J=2$; the corresponding correction to $\nu_{12}$, denoted as $\Delta_{\mathrm{mix}}$, is also listed in Table~\ref{tab:fine-structure_v12}.
To date, complete calculations of the fine-structure splittings including all terms up to order $m\alpha^7$ have been performed only for the $2\,^3P_J$ and $3\,^3P_J$ states of helium~\cite{pachucki2010, zhang2015, yan95,Mueller2005}.
For higher Rydberg states, the present work estimated the $m\alpha^6$ contributions by extrapolating the $2\,^3P_J$ results~\cite{pachucki2010} according to a $1/n^3$ scaling, following the approach adopted in previous studies~\cite{Bondy2025}. As summarized in Table~\ref{tab:fine-structure_v12}, the uncertainties of the calculated fine-structure splittings for $^4$He with principal quantum numbers $n=$ 24–37 are all below 5 kHz, indicating the high numerical precision achieved in the present C-BSBF computations.
The overall uncertainty of the fine-structure splittings is dominated by the extrapolation of the $m\alpha^6$ contribution.

As shown in Table~\ref{tab:fine-structure_v12}, comparisons are made among the present C-BSBF results, the Hylleraas calculations~\cite{drake2025}, and the QDT predictions. The fine-structure splittings for $n=$ 24–35 obtained with the Hylleraas basis were inferred from the calculated energies of the $n\,^3P_{0,1,2}$ states reported in Ref.~\cite{drake2025}. The uncertainties were determined as the square root of the sum of the squared uncertainties of the two corresponding energy levels. Since the radiative $m\alpha^6$ corrections were included in those calculations, the dominant source of uncertainty originates from the nonradiative $m\alpha^6$ contributions. QDT values listed in Table~\ref{tab:fine-structure_v12} for $n=$ 24–37 were obtained by using Drake's quantum defect parameters~\cite{drake2023springer}. QDT provides an alternative method for determining energies and fine-structure splittings of higher Rydberg states, particularly when \emph{ab initio} calculations become increasingly demanding due to electron correlation and core potential complexity~\cite{Seaton1983, Aymar1996}. As can be seen from the table, the fine-structure splittings obtained with the C-BSBFs and Hylleraas basis are consistent with each other within uncertainties, and also agree well with the QDT predictions. This cross-validation between two ab initio approaches, together with the agreement with QDT, provides an evidence for the reliability of the present C-BSBF computations.




For $n=34$–36, the present C-BSBF results enable a direct comparison between ab initio calculations and experimental measurements. The measured values of both $\nu_{12}$ and $\nu_{02}$ obtained by microwave spectroscopy~\cite{Deller2018} are listed in the last column of Table~\ref{tab:fine-structure_v12}. For $\nu_{12}$ and $\nu_{02}$, the relative deviations of the present calculations from the experimental values are below 3\% and 1.5\%, respectively. These deviations exceed the relative uncertainties of the experimental measurements, which are at the level of a few thousandths. However, the reported experimental uncertainties of 15–40 kHz do not include systematic contributions. A large deviation occurs at $n=34$, where the measured $\nu_{02}$ interval differs from the quantum-defect prediction by about 4.2$\sigma$ \cite{Deller2018}, while the present C-BSBF and Hylleraas~\cite{drake2025} results remain in excellent agreement with QDT, confirming the reliability of the theoretical framework. Better control of stray electric and magnetic fields is expected to reduce systematic uncertainties in the measured fine-structure intervals to the order of 1 kHz~\cite{Deller2018}, allowing a more stringent comparison with the ab initio theoretical calculations.


%
\begin{table}[ht]
	\caption{$1/n$ asymptotic expansion coefficients in Eq.~(\ref{operator_fit}) for the spin-dependent Breit operators and singlet-triplet mixing in the $n\,^3P_1$ states, and for the fine-structure splittings of the $n\,^3P_J$ states in $^4$He. \label{tab:coefficients}}
	\begin{threeparttable}
		\begin{tabular}{l  S[table-format=4.10] S[table-format=3.10] S[table-format=3.6] }
			\toprule
			\multirow{1}{*}{Terms}                    & \multicolumn{1}{c}{$c_0$} & \multicolumn{1}{c}{$c_1$} & \multicolumn{1}{c}{$c_2$} \\ \midrule
			$\left< H_{\mathrm{so}} \right>_1^+$      & -0.254231                 & -0.052224                 & -0.006287                 \\
			$\left< H_{\mathrm{soo}} \right>_1^+$     & 0.366653                  & 0.075264                  & 0.055483                  \\
			$\left< H_{\mathrm{ss}} \right>_1^+$      & -0.149004                 & -0.030530                 & -0.061548                 \\
			$\left< \Delta_{\mathrm{so}} \right>_1^+$ & 0.743997                  & 0.152890                  & 0.013743                  \\
			$\Delta_{\mathrm{mix}}$                   & -31.196028                & -2.609676                 & -14.351107                \\
			$\nu_{12}$                                & \num{1.596051e4}          & \num{3.277532e3}          & \num{8.577158e3}          \\
			$\nu_{02}$                                & \num{2.123714e5}          & \num{4.351912e4}          & \num{9.069520e4}          \\
			\bottomrule
		\end{tabular}
	\end{threeparttable}
\end{table}
Having calculated the expectation values of the spin-dependent operators and the singlet–triplet mixing term, as well as the fine-structure splittings for a series of $n\,^3P_J$ states, we can now analyze their $n$ dependence. The numerical results listed in Tables~\ref{tab:nonrel-energy-operator} and~\ref{tab:fine-structure_v12} were fitted to a $1/n$ asymptotic expansion of the following form:
\begin{align}
	\langle \hat{X} \rangle = \frac{1}{n^3} \sum_{i=0}^2 \frac{c_i}{n^i} \, . \label{operator_fit}
\end{align}
The obtained coefficients $c_i$ are presented in Table~\ref{tab:coefficients}. Using the parameters $c_0$, $c_1$, and $c_2$, the fitted values of $\nu_{12}$ and $\nu_{02}$ for $n=37$ are \num{0.31697} and \num{4.21720} MHz, respectively. These fitted values are in good agreement with the present ab initio fine-structure splittings, demonstrating the reliability of the $1/n$ asymptotic expansion. Using this expansion, both the contributions from the spin-dependent operators and the singlet–triplet mixing term, as well as the fine-structure splittings for higher Rydberg states, can be predicted.



%
\begin{table}[ht]
	\caption{Comparison between theoretical and experimental values of the fine-structure interval $\nu_{(0,\overline{12})}$ (in kHz) for the $n \, ^3P_J\,(n=45-51)$ states in $^4\mathrm{He}$. \label{tab:fine_structure_fit}}
	\begin{threeparttable}
		\begin{tabular}{l  S[table-format=10.8] S[table-format=7.8] S[table-format=6.2]}
			\toprule
			\multirow{1}{*}{$n \quad$} & \multicolumn{1}{c}{Present} & \multicolumn{1}{c}{QDT$\quad$} & \multicolumn{1}{c}{Exp.~\cite{Deller2018}} \\ \midrule
			45                         & 2275.656                    & 2275.66                        & 2243                                       \\
			46                         & 2130.218                    & 2130.22                        & 2150                                       \\
			47                         & 1996.914                    & 1996.92                        & 1971                                       \\
			48                         & 1874.505                    & 1874.51                        & 1817                                       \\
			49                         & 1761.899                    & 1761.90                        & 1813                                       \\
			50                         & 1658.136                    & 1658.14                        & 1646                                       \\
			51                         & 1562.363                    & 1562.36                        & 1476                                       \\
			\bottomrule
		\end{tabular}
	\end{threeparttable}
\end{table}
Using Eq.~(\ref{operator_fit}) and the fitted coefficients given in Table~\ref{tab:coefficients}, the calculated fine-structure splittings for $n=45$–51 are presented in Table~\ref{tab:fine_structure_fit}. To facilitate comparison with the experimental measurements~\cite{Deller2018}, the weighted-average values $\nu_{0,\overline{12}}$ are reported, since the fine-structure intervals decrease as $n^{-3}$, making $\nu_{12}$ increasingly difficult to resolve for higher $n$ in the experiment. The QDT predictions listed in Table~\ref{tab:fine_structure_fit} are also weight-averaged.
As shown in the table, the present fitted results are in excellent agreement with the QDT values, with discrepancies below 0.01 kHz across the entire range, demonstrating that the fitted parameters reliably describe the fine structure of the $n\,^3P$ higher Rydberg series in $^4$He. The relative deviations of the fitted results from the measurements~\cite{Deller2018} are within 6\%. This discrepancy may be attributed to unresolved theoretical effects and limitations in the experimental precision, highlighting the need for further refinement in both theory and measurement.


\section{Conclusion}

In summary, we have extended the correlated B-spline basis function (C-BSBF) method to compute the fine-structure splittings of high Rydberg states in $^4$He. The method accurately incorporates both electron–electron correlation and long-range asymptotic behavior within a unified basis, allowing stable and precise ab initio calculations for large-$n$ states. Corrections of orders $m\alpha^4$ and $m\alpha^5$, as well as singlet–triplet mixing contributions of order $m\alpha^6$ were evaluated explicitly, while the spin-dependent $m\alpha^6$ contributions were estimated using a $1/n^3$ scaling extrapolation from low-lying states.

For principal quantum numbers $n=24$–37, the present calculations achieve kHz-level accuracy in the fine-structure intervals $\nu_{12}$ and $\nu_{02}$. The results show excellent agreement with both Hylleraas-basis ab initio data and quantum-defect-theory (QDT) predictions, demonstrating the reliability of the C-BSBF approach for describing highly excited two-electron systems. Based on these high-precision results, the fine-structure splittings for $n=45$–51 were further obtained through $1/n$-asymptotic fitting, yielding values that reproduce QDT predictions within 0.01 kHz and agree with experimental measurements within 6\%.

The demonstrated capability of the C-BSBF method to achieve kHz-level precision for helium Rydberg fine structures provides a solid foundation for future inclusion of complete $m\alpha^6$ and $m\alpha^7$ QED corrections. Moreover, detailed comparisons between high-accuracy theory and precision spectroscopy of high-$n$ Rydberg states will enable stringent tests of QED and may serve as a sensitive probe for subtle or previously unaccounted physical effects in two-electron systems.

\begin{acknowledgments}
	This work is supported by the National Natural Science Foundation of China under Grants
	No. 12274417, No. 12274423, No. 12393821, No. 12204412, and No. 12174402, by the Chinese
	Academy of Sciences Project for Young Scientists in Basic Research under Grant No. YSBR-055, and by the Pioneer Research Project for Basic and Interdisciplinary Frontiers of Chinese Academy of Sciences under Grants No. XDB0920101 and XDB0920100. Theoretical calculations were done on the APM-Theoretical Computing Cluster(APM-TCC). 
\end{acknowledgments}

\section{Appendix}
This Appendix presents the fine-structure splittings of $^4$He $n\,^3P_J$ states for principal quantum numbers $11 \le n \le 37$. In addition to the present calculations, Table~\ref{tab:comparison} also includes results from quantum defect theory (QDT) and recent high-accuracy calculations by Drake \emph{et al.}~\cite{drake2025}. The uncertainties of the latter are defined as the square root of the sum of the squared uncertainties of the corresponding level energies~\cite{drake2025}. The present results are in agreement with those of Drake \emph{et al.}~\cite{drake2025} within uncertainties.

\begin{table*}[ht]
	\centering
	\caption{Comparison of fine-structure intervals between the $n \,^3P_J$ levels in $^4$He obtained from different methods (in MHz). \label{tab:comparison} \\}
	\begin{threeparttable}
		\begin{tabular*}{\textwidth}{@{\extracolsep{\fill}} c  S[table-format=4.6] S[table-format=4.6] S[table-format=2.4] S[table-format=5.6] S[table-format=5.6] S[table-format=4.7]}
			\toprule
			$n$ & \multicolumn{1}{c}{$\nu_{12}$} & \multicolumn{1}{c}{$\nu_{12}$~\cite{drake2025}} & \multicolumn{1}{c}{QDT} & \multicolumn{1}{c}{$\nu_{02}$} & \multicolumn{1}{c}{$\nu_{02}$~\cite{drake2025}} & \multicolumn{1}{c}{QDT} \\
			\midrule
			11   & 12.268(39)                     & 12.260(18)                  & 12.44             & 163.094(49)                    & 163.047(18)                &  163.27       \\
			12   & 9.429(30)                      & 9.422(14)                   & 9.50              & 125.364(38)                    & 125.328(14)                &  125.43       \\
			13   & 7.402(24)                      & 7.398(11)                   & 7.430             & 98.432(30)                     & 98.404(11)                 &  98.460       \\
			14   & 5.918(19)                      & 5.914(8)                    & 5.929 9           & 78.696(24)                     & 78.674(8)                  &  78.708 5     \\
			15   & 4.805(15)                      & 4.802(7)                    & 4.810 8           & 63.904(19)                     & 63.886(7)                  &  63.909 6     \\
			16   & 3.955(13)                      & 3.953 0(52)                 & 3.957 6           & 52.599(16)                     & 52.584 3(52)               &  52.601 9     \\
			17   & 3.294(11)                      & 3.292 6(41)                 & 3.295 4           & 43.811(13)                     & 43.799 1(41)               &  43.812 8     \\
			18   & 2.772 5(89)                    & 2.771 7(33)                 & 2.773 3           & 36.877(11)                     & 36.867 3(33)               &  36.878 2     \\
			19   & 2.355 6(75)                    & 2.355 1(27)                 & 2.356 0           & 31.333 0(95)                   & 31.324 5(27)               &  31.333 5     \\
			20   & 2.018 2(65)                    & 2.018 1(21)                 & 2.018 5           & 26.846 8(81)                   & 26.839 6(21)               &  26.847 0     \\
			21   & 1.742 4(56)                    & 1.741 9(17)                 & 1.742 5           & 23.177 8(70)                   & 23.171 7(17)               &  23.177 9     \\
			22   & 1.514 6(49)                    & 1.514 9(17)                 & 1.514 7           & 20.148 1(61)                   & 20.142 9(17)               &  20.148 2     \\
			23   & 1.324 8(43)                    & 1.325 3(16)                 & 1.324 9           & 17.624 3(53)                   & 17.619 8(16)               &  17.624 4     \\
			24   & 1.165 5(38)                    & 1.166 1(16)                 & 1.165 6           & 15.505 1(47)                   & 15.501 3(16)               &  15.505 1     \\
			25   & 1.030 7(33)                    & 1.031 5(16)                 & 1.030 8           & 13.712 5(42)                   & 13.709 2(16)               &  13.712 5     \\
			26   & 0.916 0(30)                    & 0.916 8(14)                 & 0.916 0           & 12.185 9(37)                   & 12.183 1(14)               &  12.185 9     \\
			27   & 0.817 6(26)                    & 0.818 5(14)                 & 0.817 7           & 10.877 8(33)                   & 10.875 3(14)               &  10.877 8     \\
			28   & 0.732 9(24)                    & 0.733 9(14)                 & 0.732 9           & 9.750 4(30)                    & 9.748 4(14)                &  9.750 4      \\
			29   & 0.659 5(21)                    & 0.660 5(14)                 & 0.659 5           & 8.773 6(27)                    & 8.771 8(14)                &  8.773 6      \\
			30   & 0.595 5(19)                    & 0.596 6(14)                 & 0.595 5           & 7.923 1(24)                    & 7.921 5(14)                &  7.923 1      \\
			31   & 0.539 6(17)                    & 0.540 8(13)                 & 0.539 6           & 7.179 0(22)                    & 7.177 7(13)                &  7.179 0      \\
			32   & 0.490 5(16)                    & 0.491 7(13)                 & 0.490 5           & 6.525 3(20)                    & 6.524 1(13)                &  6.525 3      \\
			33   & 0.447 1(14)                    & 0.448 3(13)                 & 0.447 1           & 5.948 6(18)                    & 5.947 6(13)                &  5.948 6      \\
			34   & 0.408 7(13)                    & 0.410 0(13)                 & 0.408 7           & 5.437 9(17)                    & 5.437 1(13)                &  5.437 9      \\
			35   & 0.374 6(12)                    & 0.375 9(13)                 & 0.374 6           & 4.984 0(15)                    & 4.983 3(13)                &  4.984 0      \\
			36   & 0.344 2(11)                    &                             & 0.344 2           & 4.579 3(14)                    &                            &  4.579 3      \\
			37   & 0.317 0(10)                    &                             & 0.317 0           & 4.217 2(13)                    &                            &  4.217 2      \\
			\bottomrule
		\end{tabular*}
	\end{threeparttable}
\end{table*}

\end{document}